\documentclass[pra,preprint,amssymb]{revtex4}
\usepackage{graphicx}
\usepackage{bm}
\usepackage{amsmath,amssymb,fancybox}
\usepackage{color}

\begin{document}
\title{Quantum Merlin-Arthur with noisy channel}
\author{Tomoyuki Morimae}
\email{morimae@gunma-u.ac.jp}
\affiliation{ASRLD Unit, Gunma University, 1-5-1 Tenjin-cho Kiryu-shi
Gunma-ken, 376-0052, Japan}
\author{Keisuke Fujii}
\email{fujii@qi.t.u-tokyo.ac.jp}
\affiliation{Photon Science Center, Graduate School of Engineering,
The University of Tokyo, 2-11-16 Yayoi, Bunkyo-ku, Tokyo 113-8656, Japan}
\author{Harumichi Nishimura}
\email{hnishimura@math.cm.is.nagoya-u.ac.jp}
\affiliation{Department of Computer Science and Mathematical
Informatics, Graduate School of Information Science,
Nagoya University, Furhocho, Chikusaku, Nagoya, Aichi, 464-8601 Japan}

\begin{abstract}
What happens if in QMA 
the quantum channel between Merlin and Arthur is noisy?
It is not difficult to show that such a modification does not change
the computational power as long as the noise
is not too strong so that errors are correctable
with high probability, 
since
if Merlin encodes
the witness state in a quantum error-correction code
and sends it to Arthur,
Arthur can correct the error caused by the noisy channel.
If we further assume that
Arthur can do only single-qubit measurements,
however, the problem becomes nontrivial,
since in this case Arthur cannot do the universal quantum computation
by himself.
In this paper,
we
show that such a restricted complexity class is still
equivalent to QMA.
To show it, we use measurement-based quantum computing: honest Merlin
sends the graph state to Arthur, and Arthur does 
fault-tolerant measurement-based quantum
computing on the noisy graph state with only single-qubit measurements. 
By measuring stabilizer operators, Arthur also checks
the correctness of the graph state. Although this idea itself was already
used in several previous papers, these results cannot be directly
used to the present case, since the test that checks the graph state
used in these papers is so strict that
even honest Merlin is rejected
with high probability
if the channel is noisy. 
We therefore introduce a more relaxed test that can accept not
only the ideal graph state but also noisy graph states that are
error-correctable.
\end{abstract}
\maketitle  

\section{Introduction}
Measurement-based quantum computing~\cite{MBQC} allows
universal quantum computing only with
adaptive single-qubit measurements on a certain
entangled state such as the graph state.
Measurement-based quantum computing has recently been applied
in quantum computational complexity theory.
For example, Ref.~\cite{Matt} used measurement-based quantum
computing to construct a multiprover
interactive proof system for BQP with a classical verifier,
and Refs.~\cite{MNS,QAMsingle} used measurement-based quantum
computing to show that the verifier needs only single-qubit measurements
in QMA and QAM.
It was also shown that the quantum state distinguishability,
which is a QSZK-complete problem, and the quantum circuit
distinguishability, which is a QIP-complete problem,
can be solved with the verifier who can do only single-qubit 
measurements~\cite{QSZKsingle}.
The basic idea in these results is the verification of the
graph state: prover(s) generate the graph state,
and the verifier performs measurement-based quantum computing on it.
By checking the stabilizer operators, the verifier
can also verify the correctness of the graph state.
We call the test ``the stabilizer test" 
(see also Refs.~\cite{HM,HF} in the context of the blind
quantum computing).
The idea of testing stabilizer operators was also used in Refs.~\cite{FV,Ji}
to construct multiprover interactive proof systems for local
Hamiltonian problems.

What happens if in QMA 
the quantum channel between Merlin and Arthur is noisy?
The first result of the present paper is 
that such a modification does not change
the computational power as long as the noise is not too
strong so that errors are correctable with high probability. 
The proof is simple: Merlin encodes
the witness state with 
a quantum error-correcting code, and sends it to Arthur
who can correct channel error by doing the
quantum error correction.

The problem becomes more nontrivial if we 
further assume that
Arthur can do only single-qubit measurements, 
since in this case Arthur cannot do the universal quantum
computation by himself.
The second result of the present paper is that the noisy QMA with
such an additional restriction for Arthur is
still equivalent to QMA.
To show it, we use measurement-based quantum computing: honest Merlin
sends the graph state to Arthur, and Arthur does 
fault-tolerant measurement-based quantum
computing on it with only single-qubit measurements. 
By measuring stabilizer operators, Arthur also checks
the correctness of the graph state. 

Note that
the results of Refs.~\cite{MNS,QAMsingle,QSZKsingle} cannot be directly
applied to the present case, since the stabilizer test 
used in these results is so strict that
even honest Merlin is rejected
with high probability
if the channel is noisy: 
even if honest Merlin sends the ideal graph state,
the state is changed due to the noise in the channel,
and such a deviated state is rejected with high probability
by the stabilizer test
in spite that the correct quantum computing
is still possible on such a state 
by correcting errors.
We therefore introduce a more relaxed test that can accept not
only the ideal graph state but also noisy graph states
that are error-correctable.
Note that recently a similar relaxed stabilizer test
was introduced and applied to blind quantum computing in
Ref.~\cite{HF}.

\section{Noisy QMA}
In this section, we define two
noisy QMA classes, 
${\rm QMA}_{\mathcal E}$
and
${\rm QMA}_{{\mathcal E},{\rm single}}$.
First we define
${\rm QMA}_{\mathcal E}$.

{\bf Definition 1}:
Let ${\mathcal E}\equiv\{\mathcal E_n\}_n$ be a family of CPTP maps,
where ${\mathcal E}_n$ is a CPTP map acting on $n$ qubits.
A language $L$ is in 
${\rm QMA}_{\mathcal E}(a,b)$ if and only if
there exists a uniformly-generated family $\{V_x\}_x$ of polynomial-size
quantum circuits such that 
\begin{itemize}
\item
If $x\in L$ then there exists an $m$-qubit
state $|\psi\rangle$ such that 
the probability of obtaining 1 when the first
qubit of 
$V_x [{\mathcal E}_m(|\psi\rangle\langle\psi|)
\otimes |0\rangle\langle0|^{\otimes n}] V_x^\dagger$ 
is measured in the computational basis
is $\ge a$. Here, $n=poly(|x|)$ and $m=poly(|x|)$.
\item
If $x\notin L$ then for any $m$-qubit state
$|\psi\rangle$, 
the probability of obtaining 1 when the first
qubit of 
$V_x [|\psi\rangle\langle\psi|
\otimes |0\rangle\langle0|^{\otimes n}] V_x^\dagger$ 
is measured in the computational basis
is $\le b$.
\end{itemize}

Note that this definition reflects a physically natural assumption
that malicious Merlin can replace the channel, and therefore
Arthur should assume that any state can be sent in no cases.
We can also consider another definition that assumes
that even evil Merlin cannot modify the
channel, but in this case we do not know how to show
that the class is in QMA, and therefore in this paper,
we do not consider the definition.

We can show that ${\rm QMA}_{\mathcal E}$ contains ${\rm QMA}$
if ${\mathcal E}$ is not too strong so that errors are
correctable with high probability.
(More details about the error correctability is 
given in Sec.~\ref{sec:fujii}.)
Throughout this paper, we assume that ${\mathcal E}$
satisfies such property, since if the channel noise is
too strong and therefore the witness state is
completely destroyed,
the noisy QMA is trivially in BQP.

{\bf Theorem 1}:
For any $(a,b)$ such that $a-b\ge 1/poly(|x|)$ 
and any $r=poly(|x|)$,
\begin{eqnarray*}
{\rm QMA}(a,b)\subseteq{\rm QMA}_{\mathcal E}(1-2^{-r},2^{-r}).
\end{eqnarray*}

{\bf Proof}:
Let us assume that a language $L$ is in 
${\rm QMA}(a,b)$. 
Then, there exists a uniformly-generated family $\{V_x\}_x$
of polynomial-size quantum circuits such that
\begin{itemize}
\item
If $x\in L$ then there exists an $m$ qubit state
$|\psi\rangle$ such that the probability of
obtaining 1 when the first qubit of
$V_x(|\psi\rangle\otimes|0\rangle^{\otimes n})$ 
is measured in the computational basis is
$\ge a$, where $n=poly(|x|)$ and $m=poly(|x|)$.
\item
If $x\notin L$ then for any $m$ qubit state
$|\psi\rangle$, the probability
is
$\le b$.
\end{itemize}

According to the standard argument of
the error reduction, for any
polynomial $t$,
there exists a uniformly-generated family
$\{V_x'\}_x$ of polynomial-size quantum circuits such that
\begin{itemize}
\item
If $x\in L$ then the probability of obtaining 1 when
the first qubit of 
$V_x'(|\psi\rangle^{\otimes k}\otimes|0\rangle^{\otimes n'})$ is measured
in the computational basis is
$\ge 1-2^{-t(|x|)}$,
where $k=poly(|x|)$ and $n'=poly(|x|)$.
\item
if $x\notin L$ then for any $mk$ qubit state,
the probability is
$\le 2^{-t(|x|)}$.
\end{itemize}

From $V_x'$, we construct the circuit $V_x''$ that first
does the error correction and decoding, and then
applies $V_x'$.
If $x\in L$, honest Merlin sends Arthur 
$Enc(|\psi\rangle^{\otimes k})$, which is
the encoded version of $|\psi\rangle^{\otimes k}$
in a certain quantum error-correcting code.
Due to the noise, what Arthur receives is
${\mathcal E}_u(Enc(|\psi\rangle^{\otimes k}))$, where $u$ is the size
of $Enc(|\psi\rangle^{\otimes k})$.
By definition, errors are correctable,
and therefore,
according to the theory of quantum error correction~\cite{NC},
for any polynomial $s$, there exists a number of
the repetitions of the concatenation
such that $u=poly(m)$ and
the state $\rho$ after the error correction and decoding
on ${\mathcal E}_u(Enc(|\psi\rangle^{\otimes k}))$
satisfies
\begin{eqnarray*}
\frac{1}{2}\Big\|\rho-|\psi\rangle\langle\psi|^{\otimes k}\Big\|_1\le
2^{-s}.
\end{eqnarray*}
If $V_x'$ is applied on $\rho$,
the acceptance probability is
\begin{eqnarray*}
p_{acc}&\ge& (1-2^{-t})-2^{-s}\\
&\ge& 1-2^{-r},
\end{eqnarray*}
where we have taken sufficiently large $k$ and the number
of the repetitions of the concatenation such that
\begin{eqnarray*}
2^{-s}&\le& 2^{-r-1},\\
2^{-t}&\le& 2^{-r-1}.
\end{eqnarray*}
Therefore, the probability that $V_x''$ accepts
${\mathcal E}_u(Enc(|\psi\rangle^{\otimes k}))$ 
is 
larger than $1-2^{-r}$.
If $x\notin L$, on the other hand, any state is
accepted by $V_x'$ with probability at most $2^{-t}$.
It is also the case for the output of
the error-correcting and decoding circuit on any input.
Therefore, the acceptance probability of $V_x''$ on any state is
\begin{eqnarray*}
p_{acc}&\le& 2^{-t}\\
&\le& 2^{-r-1}\\
&\le& 2^{-r}.
\end{eqnarray*}
Hence we have shown that the language $L$ is in
${\rm QMA}_{\mathcal E}(1-2^{-r},2^{-r})$.
$\blacksquare$

We next define the class
${\rm QMA}_{{\mathcal E},{\rm single}}(a,b)$.

{\bf Definition 2}:
The class ${\rm QMA}_{{\mathcal E},{\rm single}}(a,b)$
is the restricted version of
${\rm QMA}_{{\mathcal E}}(a,b)$
such that Arthur can do only single-qubit measurements.

Our second result is the following theorem.

{\bf Theorem 2}:
For any $(a,b)$ such that $a-b\ge1/poly(|x|)$
and any $r=poly(|x|)$,
\begin{eqnarray*}
{\rm QMA}(a,b)
\subseteq{\rm QMA}_{{\mathcal E},{\rm single}}(1-2^{-r},2^{-r}).
\end{eqnarray*}

The rest of the paper is devoted to show Theorem 2.

\section{Measurement-based quantum computing}
\label{Sec:MBQC}
For readers unfamiliar with measurement-based quantum
computing, we here explain some basics.
Let us consider a graph $G=(V,E)$, 
where $|V|=N$.
The graph state $|G\rangle$ on $G$ is defined by
\begin{eqnarray*}
|G\rangle\equiv
\Big(\prod_{(i,j)\in E}CZ_{i,j}\Big)
|+\rangle^{\otimes N},
\end{eqnarray*}
where $|+\rangle\equiv(|0\rangle+|1\rangle)/\sqrt{2}$, and
$CZ_{i,j}\equiv|0\rangle\langle0|\otimes I+|1\rangle\langle1|\otimes Z$ 
is the $CZ$ gate on the vertices $i$ and $j$.

According to the theory of measurement-based quantum computing~\cite{MBQC},
for any $m$-width $d$-depth quantum circuit $U$,
there exists a graph $G=(V,E)$ for $|V|=N=poly(m,d)$
and the graph state $|G\rangle$
on it such that if we measure each qubit 
in $V-V_o$, where $V_o$ is a certain
subset of $V$ with $|V_o|=m$, in certain bases adaptively,
then the state of $V_o$ after the measurements
is
\begin{eqnarray*}
B_{x,z}^mU|0^m\rangle
\end{eqnarray*}
with uniformly randomly chosen 
$x\equiv(x_1,...,x_m)\in\{0,1\}^m$ 
and $z\equiv(z_1,...,z_m)\in\{0,1\}^m$,
where 
\begin{eqnarray*}
B_{x,z}^m\equiv\bigotimes_{j=1}^m
X_j^{x_j}Z_j^{z_j}.
\end{eqnarray*}
The operator is called a byproduct operator,
and its effect is corrected, since $x$ and $z$ can be
calculated from previous measurement results.
Hence we finally obtain the desired state $U|0^m\rangle$.

If we entangle each qubit of a state $|\psi\rangle$ with
an appropriate qubit of $|G\rangle$ by using $CZ$ gate,
we can also implement $U|\psi\rangle$ in measurement-based quantum
computing.

The graph state $|G\rangle$ is stabilized by
\begin{eqnarray}
g_j\equiv X_j\bigotimes_{i\in S_j}Z_i,
\label{stabilizer}
\end{eqnarray}
for all $j\in V$, where $S_j$ is the set of
nearest-neighbour vertices of $j$th vertex. 
In other words,
\begin{eqnarray*}
g_j|G\rangle=|G\rangle
\end{eqnarray*}
for all $j\in V$. 

For $u\equiv(u_1,...,u_N)\in\{0,1\}^N$,
we define the state $|G_u\rangle$
by
\begin{eqnarray*}
g_j|G_u\rangle=(-1)^{u_j}|G_u\rangle
\end{eqnarray*}
for all $j\in V$.
(Therefore, $|G\rangle=|G_{0^N}\rangle$.)
The set $\{|G_u\rangle\}_u$ is an
orthonormal basis of the $N$-qubit Hilbert space.
In fact, if $u\neq u'$, there exists $j$ such that $u_j\neq u_j'$.
Then,
\begin{eqnarray*}
\langle G_{u'}|G_u\rangle&=&
\langle G_{u'}|g_jg_j|G_u\rangle\\
&=& (-1)^{u_j+u_j'}\langle G_{u'}|G_u\rangle\\
&=& -\langle G_{u'}|G_u\rangle,
\end{eqnarray*}
and therefore $\langle G_{u'}|G_u\rangle=0$.

\section{Stabilizer test}
\label{Sec:ST}
\begin{figure}[htbp]
\begin{center}
\includegraphics[width=0.4\textwidth]{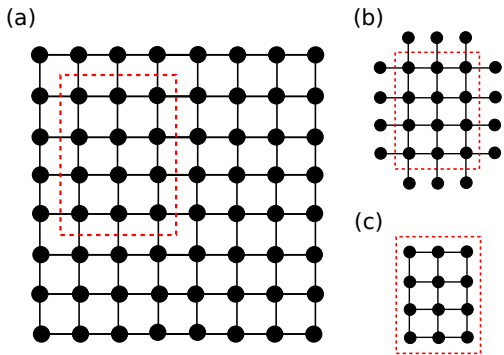}
\end{center}
\caption{
(a) The graph $G$. $V_1$ is the set of vertices in the
dotted red square, and  $V_2$ is the set of other vertices.
(b) The subgraph $G'$.
(c) The subgraph $G''$.
} 
\label{stabilizer_test}
\end{figure}

For the convenience of readers,
we also review the stabilizer test used in 
Refs.~\cite{MNS,QAMsingle,QSZKsingle}.
Consider the graph $G=(V,E)$ of Fig.~\ref{stabilizer_test}.
(For simplicity, we here consider the square lattice,
but the result can be applied to any reasonable graph.)
As is shown in Fig.~\ref{stabilizer_test}, 
we define two subsets, $V_1$ and $V_2\equiv V-V_1$, of $V$,
where $|V_1|=N_1$ and $|V_2|=N_2$.
We also define a subset $V_{connect}$ of $V_2$ by
\begin{eqnarray*}
V_{connect}\equiv\{j\in V_2|\exists i\in V_1~\mbox{s.t}~(i,j)\in E
\}.
\end{eqnarray*}
In other words, $V_{connect}$ is the set of vertices in $V_2$ that are
connected to vertices in $V_1$.
We further define two subsets of $E$:
\begin{eqnarray*}
E_1&\equiv&\{(i,j)\in E|i\in V_1~\mbox{and}~j\in V_1\},\\
E_{connect}&\equiv&\{(i,j)\in E|i\in V_1~\mbox{and}~j\in V_2\}.
\end{eqnarray*}
Finally, we define two subgraphs
of $G$:
\begin{eqnarray*}
G'&\equiv&(V_1\cup V_{connect},E_1\cup E_{connect}),\\
G''&\equiv&(V_1,E_1).
\end{eqnarray*}

The stabilizer test is the following test:
\begin{itemize}
\item[1.]
Randomly generate an $N_1$-bit string 
$k\equiv(k_1,...,k_{N_1})\in\{0,1\}^{N_1}$.
\item[2.]
Measure the operator
\begin{eqnarray*}
s_k\equiv\prod_{j\in V_1}(g_j')^{k_j},
\end{eqnarray*}
where
$g_j'$ is the stabilizer operator, Eq.~(\ref{stabilizer}), of the graph state
$|G'\rangle$.
\item[3.]
If the result is $+1$ $(-1)$, the test passes (fails).
\end{itemize}
Let $|\Psi\rangle$ be a pure state on $V$.
If the probability $p_{test}$ that $|\Psi\rangle$
passes the stabilizer test
satisfies $p_{test}\ge1-\epsilon$, where 
$0\le \epsilon\le \frac{1}{2}$, then
\begin{eqnarray*}
\frac{1}{2}
\Big\||\Psi\rangle\langle\Psi|
-|\Psi'\rangle\langle\Psi'|
\Big\|_1\le\sqrt{4\epsilon-4\epsilon^2},
\end{eqnarray*}
where 
\begin{eqnarray*}
|\Psi'\rangle\equiv
W(|G''\rangle\otimes|\xi\rangle_{V_2}).
\end{eqnarray*}
Here, $|\xi\rangle$ is a certain state on $V_2$
and
\begin{eqnarray*}
W\equiv\prod_{(i,j)\in E_{connect}}CZ_{i,j}.
\end{eqnarray*}
For a proof, see Ref.~\cite{QSZKsingle}.

\section{Necessity of more relaxed test}
According to the theory of fault-tolerant measurement-based 
quantum computing,
if ${\mathcal E}$ is not too strong,
fault-tolerant measurement-based quantum computing is possible
on the state ${\mathcal E}_n(|G\rangle\langle G|)$ 
for a certain $n$-qubit graph $G$~\cite{RHG}.
In particular, there exists a set 
$\Gamma\subset\{0,1\}^n$ 
of $n$-bit strings $\gamma$ such that 
fault-tolerant measurement-based quantum computing is possible
on $|G_\gamma\rangle$.
(For more details, see Sec.~\ref{sec:fujii}.)

If there is some noise in the quantum channel between
Merlin and Arthur,
the stabilizer test introduced in the previous section
is so strict that even honest Merlin is rejected with high probability.
For example, let
us assume that honest Merlin sends Arthur the correct state
\begin{eqnarray*}
W(|G''\rangle\otimes|\xi\rangle_{V_2}),
\end{eqnarray*}
but due to the noise, what Arthur receives is
\begin{eqnarray*}
|\Psi\rangle=W(|G_\gamma''\rangle\otimes|\xi\rangle_{V_2})
\end{eqnarray*}
where $\gamma\in\Gamma$ but $\gamma\neq 0^{N_1}$. 
Here, 
$\Gamma\subset\{0,1\}^{N_1}$ 
is the set of $N_1$-bit strings $\gamma$ such that 
fault-tolerant measurement-based quantum computing is possible
on $|G_\gamma''\rangle$. (See Sec.~\ref{sec:fujii}.)
Then, the probability $p_{test}$ 
that $|\Psi\rangle$ passes the stabilizer test
is
\begin{eqnarray*}
p_{test}&=&\frac{1}{2^{N_1}}\sum_{k\in\{0,1\}^{N_1}}
\langle\Psi|\frac{I+s_k}{2}|\Psi\rangle\\
&=&\frac{1}{2}+\frac{1}{2}
\langle\Psi|\prod_{j\in V_1}\frac{I+g_j'}{2}|\Psi\rangle\\
&=&\frac{1}{2}.
\end{eqnarray*}
Note that this value $1/2$ is the minimum value of
$p_{test}$, since
\begin{eqnarray*}
\langle\Phi|\prod_{j\in V_1}\frac{I+g_j'}{2}|\Phi\rangle
\ge0
\end{eqnarray*}
for any state $|\Phi\rangle$.

Let us try to prove Theorem 2 by 
using the stabilizer test of the previous section.
We first assume that a language $L$ is in ${\rm QMA}(a,b)$.
Due to the error reducibility of QMA, the assumption
$L\in{\rm QMA}(a,b)$ means
that $L$ is in ${\rm QMA}(1-2^{-t},2^{-t})$
for any polynomial $t$.
We want to show that $L$ is in 
${\rm QMA}_{{\mathcal E},{\rm single}}(1-2^{-r},2^{-r})$
for any $r=poly(|x|)$.
To show it, we consider a similar protocol of Ref.~\cite{MNS}
where Arthur chooses the computation with probability $q$
and the stabilizer test with probability $1-q$.
Let $p_{comp}$ be the probability of accepting the computation
result when he chooses the computation,
and $p_{test}$ be that of passing the stabilizer test when
he chooses the stabilizer test.

First let us consider the case of $x\in L$.
In this case, Merlin sends the correct state,
i.e., the encoded witness state entangled with the graph state.
According to the theory of fault-tolerant measurement-based
quantum computing,
Arthur can do the correct quantum computing
on the noisy graph state
with probability $1-2^{-s}$ and fails the correct computing
with probability $2^{-s}$ for any polynomial $s$.
The acceptance probability $p_{acc}$ is
therefore
\begin{eqnarray*}
p_{acc}&=&qp_{comp}+(1-q)p_{test}\\
&\ge&q[(1-2^{-t})(1-2^{-s})+0\times 2^{-s}]+
(1-q)\frac{1}{2}\\
&=&q[(1-2^{-t})(1-2^{-s})]+
(1-q)\frac{1}{2}
\equiv\alpha.
\end{eqnarray*}

Next let us consider the case of $x\notin L$. 
In this case, the acceptance probability is
\begin{eqnarray*}
p_{acc}
&=&qp_{comp}+(1-q)p_{test}\\
&\le&q\times 1+(1-q)(1-\epsilon)\\
&=&q+(1-q)(1-\epsilon)
\equiv\beta
\end{eqnarray*}
if malicious Merlin sends a state such that $p_{test}<1-\epsilon$.
The gap $\Delta$ is then
\begin{eqnarray*}
\Delta\equiv\alpha-\beta=
q[(1-2^{-t})(1-2^{-s})-1]+(1-q)\Big(\epsilon-\frac{1}{2}\Big)\le 0
\end{eqnarray*}
for any $0\le q\le 1$,
and therefore we cannot show 
$\Delta\ge 1/poly$, which is necessary
to show Theorem 2.

A reason why the above proof does not work
is that the probability that honest Merlin passes the stabilizer
test is too small.
If Merlin is honest and if the channel gives only a weak
error that is correctable, what Arthur receives should be accepted with
high probability, since it is useful
for the correct quantum computing.
This argument suggests that the stabilizer test in the previous
section is too strict for several practical situations such as
the noisy channel case. 
Hence we need a more relaxed test.

\section{Proof of Theorem 2}
\begin{figure}[htbp]
\begin{center}
\includegraphics[width=0.15\textwidth]{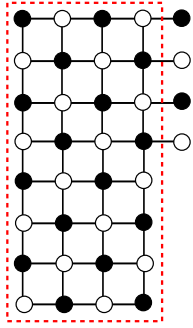}
\end{center}
\caption{
The graph $G$. $V_1$ is the set of vertices in the
dotted red square, and  $V_2$ is the set of other vertices.
Two subgraphs $G'$ and $G''$ are defined as in 
Sec.~\ref{Sec:ST}.
In this example, the subgraph $G'$ is equal to $G$.  
} 
\label{protocol}
\end{figure}

Now we give a proof of Theorem 2 by introducing a more
relaxed stabilizer test.
Let us assume that a language $L$ is in ${\rm QMA}(a,b)$.
Due to the error reducibility of QMA, this means that
$L$ is in ${\rm QMA}(1-2^{-t},2^{-t})$ for any 
polynomial $t$.
Therefore, without loss of generality, we take $a=1-2^{-t}$
and $b=2^{-t}$ for any polynomial $t$.
Let $\{V_x\}_x$ be Arthur's verification circuits,
and $|\psi\rangle$ be the yes witness that gives
the acceptance probability larger than $a=1-2^{-t}$.
We consider the bipartite graph $G$ of Fig.~\ref{protocol}.
(For simplicity, the graph is written as
the two-dimensional square lattice, 
but the graph can be more complicated depending on the computation.)

Our protocol runs as follows.
\begin{itemize}
\item[1.]
If Merlin is honest, he generates the correct state
\begin{eqnarray*}
|\Psi_{correct}\rangle\equiv
W[|G''\rangle\otimes Enc(|\psi\rangle)]
\end{eqnarray*}
on the graph $G$, where 
$Enc(|\psi\rangle)$ is the encoded version
of $|\psi\rangle$ and placed on $V_2$.
Merlin sends each qubit of $|\Psi_{correct}\rangle$ one by one to Arthur.
If Merlin is malicious, he generates any state $|\Psi\rangle$ 
on $G$ and sends each qubit of it one by one to Arthur.
(Due to the convexity, we can assume without loss of generality
that malicious Merlin sends pure states.)
\item[2.]
With probability $q$, which will be specified later,
Arthur does the fault-tolerant measurement-based
quantum computation that implements the fault-tolerant
version of $V_x$ with input $|\psi\rangle$. 
If the result is accept (reject), he accepts (rejects).
We denote the acceptance probability by $p_{comp}$.
\item[3.]
With probability $\frac{1-q}{2}$, Arthur measures
all black qubits of $G''$ in $X$ and all white qubits of $G'$ in $Z$.
Let $\{x_j\}_j$ and $\{z_j\}_j$ be the set of the $X$ measurement
results and $Z$ measurement results, respectively.
If and only if the syndrome set
\begin{eqnarray*}
Synd_1\equiv
\Big\{x_j\oplus\bigoplus_{i\in S_j}z_i\Big\}_{j\in V_1^b}
\end{eqnarray*}
satisfies certain condition $Cond_1$,
which will be explained later, 
Arthur accepts. Here, $S_j$ is the set of the nearest-neighbour
vertices of $j$th vertex in terms of the graph $G'$,
and $V_1^b$ is the set of black vertices in $V_1$.
We denote the acceptance probability by $p_{test1}$.
\item[4.]
With probability $\frac{1-q}{2}$, Arthur measures
all white qubits of $G''$ in $X$ and all black qubits of $G'$ in $Z$.
Let $\{x_j\}_j$ and $\{z_j\}_j$ be the set of the $X$ measurement
results and $Z$ measurement results, respectively.
If and only if the syndrome set
\begin{eqnarray*}
Synd_2\equiv
\Big\{x_j\oplus\bigoplus_{i\in S_j}z_i\Big\}_{j\in V_1^w}
\end{eqnarray*}
satisfies certain condition $Cond_2$,
which will be explained later, 
Arthur accepts. Here, $V_1^w$ is the set of the white
vertices in $V_1$.
We denote the acceptance probability by $p_{test2}$.
\end{itemize}
The conditions $Cond_1$ and $Cond_2$ are taken in such a way that
if $Synd_1$ satisfies $Cond_1$
and $Synd_2$ satisfies $Cond_2$
then errors 
are correctable, and therefore fault-tolerant 
measurement-based quantum computing is possible.
In this paper, we do not give the explicit expressions
of $Cond_1$ and $Cond_2$, since they are complicated and 
not necessary. 
At least, according to the theory of fault-tolerant quantum
computing, we can define such $Cond_1$ and $Cond_2$,
and the membership of $Cond_1$ and $Cond_2$ 
can be decided in a polynomial time.
A more detailed discussion is given in Sec.~\ref{sec:fujii}.

First we consider the case when
$x\in L$. Since ${\mathcal E}$ is not too strong,
$p_{test1}\ge1-\delta$ and 
$p_{test2}\ge1-\delta$ for certain 
$\delta=2^{-poly}$(see Sec.~\ref{sec:fujii}).
Therefore, the acceptance probability $p_{acc}$ is
\begin{eqnarray*}
p_{acc}&=&qp_{comp}+\frac{1-q}{2}p_{test1}+\frac{1-q}{2}p_{test2}\\
&\ge&q[(1-2^{-s})a+2^{-s}\times0]
+\frac{1-q}{2}(1-\delta)+\frac{1-q}{2}(1-\delta)\\
&=&q(1-2^{-s})a+(1-q)(1-\delta)\equiv\alpha.
\end{eqnarray*}

Next we consider the case when $x\notin L$.
There are four
possibilities for $|\Psi\rangle$:
\begin{itemize}
\item[1.]
$p_{test1}\ge 1-\epsilon$~\mbox{and}~$p_{test2}< 1-\epsilon$,
\item[2.]
$p_{test1}< 1-\epsilon$~\mbox{and}~$p_{test2}\ge 1-\epsilon$,
\item[3.]
$p_{test1}< 1-\epsilon$~\mbox{and}~$p_{test2}< 1-\epsilon$,
\item[4.]
$p_{test1}\ge 1-\epsilon$~\mbox{and}~$p_{test2}\ge 1-\epsilon$,
\end{itemize}
where $\epsilon$ is a certain parameter that will be specified
later.
Let us consider each case separately.
First, if $p_{test1}\ge 1-\epsilon$ and $p_{test2}< 1-\epsilon$,
\begin{eqnarray*}
p_{acc}&=&qp_{comp}+\frac{1-q}{2}p_{test1}+\frac{1-q}{2}p_{test2}\\
&<&q+\frac{1-q}{2}+\frac{1-q}{2}(1-\epsilon)\equiv\beta_1.
\end{eqnarray*}
Second, if $p_{test1}< 1-\epsilon$ and $p_{test2}\ge 1-\epsilon$,
\begin{eqnarray*}
p_{acc}&=&qp_{comp}+\frac{1-q}{2}p_{test1}+\frac{1-q}{2}p_{test2}\\
&<&q+\frac{1-q}{2}(1-\epsilon)+\frac{1-q}{2}=\beta_1.
\end{eqnarray*}
Third, if $p_{test1}< 1-\epsilon$ and $p_{test2}< 1-\epsilon$,
\begin{eqnarray*}
p_{acc}&=&qp_{comp}+\frac{1-q}{2}p_{test1}+\frac{1-q}{2}p_{test2}\\
&<&q+\frac{1-q}{2}(1-\epsilon)+\frac{1-q}{2}(1-\epsilon)\\
&=&q+(1-q)(1-\epsilon)\equiv\beta_2.
\end{eqnarray*}
Finally, if $p_{test1}\ge 1-\epsilon$ and $p_{test2}\ge 1-\epsilon$,
\begin{eqnarray*}
p_{acc}&=&qp_{comp}+\frac{1-q}{2}p_{test1}+\frac{1-q}{2}p_{test2}\\
&\le&q[(1-2^{-s})b+2^{-s}\times 1
+\sqrt{4\epsilon-4\epsilon^2}]+\frac{1-q}{2}+\frac{1-q}{2}\\
&=&q[(1-2^{-s})b+2^{-s}+\sqrt{4\epsilon-4\epsilon^2}]+1-q\equiv\beta_3.
\end{eqnarray*}
Here, we have used the fact that
if 
$p_{test1}\ge1-\epsilon$ and $p_{test2}\ge1-\epsilon$ then
\begin{eqnarray}
\frac{1}{2}\Big\|
|\Psi\rangle-|\Psi'\rangle\Big\|_1
\le\sqrt{4\epsilon-4\epsilon^2},
\label{noisy_stabilizer}
\end{eqnarray}
where
\begin{eqnarray*}
|\Psi'\rangle\equiv\sum_{\gamma\in\Gamma}\sum_t
D_{\gamma,t}W(|G_\gamma''\rangle\otimes|\phi_t\rangle),
\end{eqnarray*}
$\Gamma$ is the set of $\gamma$ such that
errors on $|G_\gamma''\rangle$
are correctable,
$\{D_{\gamma,t}\}_{\gamma,t}$ 
is the set of certain complex coefficients
such that 
\begin{eqnarray*}
\sum_{\gamma\in\Gamma}\sum_t|D_{\gamma,t}|^2=1,
\end{eqnarray*}
and $\{\phi_t\}_t$ is an orthonormal basis on $V_2$.
A proof of Eq.~(\ref{noisy_stabilizer}) is given in the next section.

Let us define
\begin{eqnarray*}
\Delta_1(q)&\equiv&\alpha-\beta_1=q[(1-2^{-s})a-1]+(1-q)
\Big(\frac{\epsilon}{2}-\delta\Big),\\
\Delta_2(q)&\equiv&\alpha-\beta_2=q[(1-2^{-s})a-1]+(1-q)(\epsilon-\delta),\\
\Delta_3(q)&\equiv&\alpha-\beta_3=
q[(1-2^{-s})(a-b)-2^{-s}-\sqrt{4\epsilon-4\epsilon^2}]-(1-q)\delta.
\end{eqnarray*}
Then, the value $q^*$ that gives
$\max_q\min(\Delta_1,\Delta_2,\Delta_3)$
is $q$ such that $\Delta_1(q)=\Delta_3(q)$. Therefore,
\begin{eqnarray*}
q^*\equiv\frac{\frac{\epsilon}{2}}
{1+\frac{\epsilon}{2}-(1-2^{-s})b-2^{-s}-\sqrt{4\epsilon-4\epsilon^2}}
\end{eqnarray*}
and for this $q^*$, the gap is
\begin{eqnarray*}
\Delta_3(q^*)
&=&
\frac{\frac{\epsilon}{2}[(1-2^{-s})(a-b)-2^{-s}-\sqrt{4\epsilon-4\epsilon^2}]}
{1+\frac{\epsilon}{2}-(1-2^{-s})b-2^{-s}-\sqrt{4\epsilon-4\epsilon^2}}
-\delta(1-q^*)\\
&\ge&
\frac{\frac{\epsilon}{2}[(1-2^{-s})(a-b)-2^{-s}-\sqrt{4\epsilon}]}
{1+\frac{\epsilon}{2}}
-\delta\times1\\
&=&
\frac{\frac{\epsilon}{2}[(1-2^{-s})(1-2^{-t+1})-2^{-s}-\sqrt{4\epsilon}]}
{1+\frac{\epsilon}{2}}
-\delta\\
&\ge&
\frac{\frac{1}{64\times2}(\frac{7}{8}\times\frac{7}{8}-\frac{1}{8}
-\frac{2}{8})}
{1+\frac{1}{64\times2}}
-\delta\\
&=&\frac{25}{64^2\times2+64}-\delta,
\end{eqnarray*}
where we have taken $\epsilon=\frac{1}{64}$,
$s\ge3$, and $t\ge 4$.
Hence $L$ is in 
${\rm QMA}_{{\mathcal E},{\rm single}}(a',b')$ with
$a'-b'\ge const.\ge 1/poly(|x|)$.

It is easy to show that 
if we run the above protocol in parallel,
and Arthur
takes the majority voting, then the error $(a',b')$ can be
amplified to $(1-2^{-r},2^{-r})$ for any $r=poly(|x|)$.
The proof is almost the same as that of the standard error reduction
in QMA. One different point is, however, 
that 
when the channel is noisy, even the yes witness is not the tensor product of
the original witness states, because the noise can generate
entanglement among them. 
This means that unlike the standard QMA case,
the output of each run is not independent even in the
yes case, and therefore the Chernoff bound
does not seem to be directly used. However, we can show
that the probability of obtaining 0 in the $i$th run
is upperbounded by $1-a$ whatever results obtained
in the previous runs. Therefore, the rejection probability
is upperbounded by that of the case when each run is
the independent Bernoulli trial with the coin bias $(1-a,a)$,
where the standard Chernoff bound argument works.
(More precisely, the argument is as follows.
In the first run, the probability of obtaining 0
is $Pr[y_1=0]\le 1-a$, where $y_1$ is the result of the first run.
If we assume $Pr[y_1=0]=1-a$, we can maximize the rejection probability.
In the second run, the probability of obtaining 0
is $Pr[y_2=0|y_1]\le 1-a$. If we assume
$Pr[y_2=0|y_1]=1-a$, we can maximize the rejection probability.
If we repeat it for all runs, we conclude that the
the independent Bernoulli trial with the coin bias $(1-a,a)$
achieves the maximum rejection probability.
According to the Chernoff bound, the maximum rejection probability
is upperbounded by an exponentially decaying function.
)
$\blacksquare$

\section{Proof of Eq.~(\ref{noisy_stabilizer})}
In this section, we show Eq.~(\ref{noisy_stabilizer}).
Let us define
$N_1^b\equiv|V_1^b|$
and $N_1^w\equiv|V_1^w|$.
Since $p_{test1}\ge1-\epsilon$,
\begin{eqnarray}
p_{test1}=\sum_{\omega\in\Omega_1}\langle\Psi|\prod_{j\in V_1^b}
\frac{I+(-1)^{\omega_j}g_j'}{2}|\Psi\rangle\ge1-\epsilon,
\label{assumption}
\end{eqnarray}
where $\Omega_1$ is the set of $N_1^b$-bit string
$\omega=\{\omega_j\}_{j\in V_1^b}\in\{0,1\}^{N_1^b}$ such that
$\omega$ satisfies $Cond_1$,
and $g_j'$ is the stabilizer operator of $|G'\rangle$ on $j$th qubit.
Since 
\begin{eqnarray*}
\Big\{W(|G_{(u,v)}''\rangle\otimes|\phi_t\rangle)\Big\}
_{u\in\{0,1\}^{N_1^b},v\in\{0,1\}^{N_1^w},t\in\{0,1\}^{N_2}}
\end{eqnarray*}
is an orthonormal basis,
we can write
\begin{eqnarray*}
|\Psi\rangle=
\sum_{u\in\{0,1\}^{N_1^b}}
\sum_{v\in\{0,1\}^{N_1^w}}
\sum_{t\in\{0,1\}^{N_2}}
C_{u,v,t}W(|G_{(u,v)}''\rangle\otimes|\phi_t\rangle),
\end{eqnarray*}
with certain complex coefficients $\{C_{u,v,t}\}_{u,v,t}$
such that $\sum_{u,v,t}|C_{u,v,t}|^2=1$. 
Let $\{g_j''\}_j$ be the set of stabilizer operators of the graph state
$|G''\rangle$.
Then, it is easy to check
\begin{eqnarray*}
g_j' W=Wg_j''
\end{eqnarray*}
for all $j\in V_1$.
Therefore, from Eq.~(\ref{assumption}),
\begin{eqnarray*}
1-\epsilon&\le&
\sum_{\omega\in\Omega_1}\langle\Psi|
\prod_{j\in V_1^b}
\frac{I+(-1)^{\omega_j}g_j'}{2}
|\Psi\rangle\\
&=&
\sum_{\omega\in\Omega_1}\langle\Psi|
\prod_{j\in V_1^b}
\frac{I+(-1)^{\omega_j}g_j'}{2}
\sum_{u,v,t}C_{u,v,t}W(|G_{(u,v)}''\rangle\otimes|\phi_t\rangle)\\
&=&\sum_{\omega\in\Omega_1}\langle\Psi|
W
\prod_{j\in V_1^b}
\frac{I+(-1)^{\omega_j}g_j''}{2}
\sum_{u,v,t}C_{u,v,t}(|G_{(u,v)}''\rangle\otimes|\phi_t\rangle)\\
&=&
\sum_{\omega\in\Omega_1}\langle\Psi|
W\sum_v\sum_tC_{\omega,v,t}
(|G_{(\omega,v)}''\rangle\otimes|\phi_t\rangle)\\
&=&
\sum_{\omega\in\Omega_1}\sum_v
\sum_t
|C_{\omega,v,t}|^2.
\end{eqnarray*}
In a similar way, $p_{test2}\ge1-\epsilon$ leads to
\begin{eqnarray*}
\sum_u
\sum_{\omega'\in\Omega_2}
\sum_t
|C_{u,\omega',t}|^2\ge 1-\epsilon,
\end{eqnarray*}
where $\Omega_2$ is the set of $\omega$ that
satisfy $Cond_2$.

Let us define
\begin{eqnarray*}
|\Psi'\rangle\equiv\frac{1}{\sqrt{R}}
\sum_{u\in\Omega_1}
\sum_{v\in\Omega_2}
\sum_t
C_{u,v,t}W(|G_{(u,v)}''\rangle\otimes|\phi_t\rangle)
\end{eqnarray*}
where 
\begin{eqnarray*}
R\equiv
\sum_{u\in\Omega_1}
\sum_{v\in\Omega_2}
\sum_t
|C_{u,v,t}|^2\le1
\end{eqnarray*}
is the normalization constant.
Then,
\begin{eqnarray*}
\langle\Psi|\Psi'\rangle&=&\frac{1}{\sqrt{R}}
\sum_{u\in\Omega_1}
\sum_{v\in\Omega_2}
\sum_t|C_{u,v,t}|^2\\
&\ge&
\sum_{u\in\Omega_1}
\sum_{v\in\Omega_2}
\sum_t|C_{u,v,t}|^2\\
&\ge&1-2\epsilon.
\end{eqnarray*}
Here, in the last inequality, we have used the relation
\begin{eqnarray*}
YY&\ge&YY-NN\\
&=&YY-(1-YY-YN-NY)\\
&=&(YY+YN)+(YY+NY)-1\\
&\ge&(1-\epsilon)+(1-\epsilon)-1\\
&=&1-2\epsilon,
\end{eqnarray*}
where
\begin{eqnarray*}
YY&=&\sum_{u\in\Omega_1}\sum_{v\in\Omega_2}\sum_t|C_{u,v,t}|^2,\\
YN&=&\sum_{u\in\Omega_1}\sum_{v\notin\Omega_2}\sum_t|C_{u,v,t}|^2,\\
NY&=&\sum_{u\notin\Omega_1}\sum_{v\in\Omega_2}\sum_t|C_{u,v,t}|^2,\\
NN&=&\sum_{u\notin\Omega_1}\sum_{v\notin\Omega_2}\sum_t|C_{u,v,t}|^2.
\end{eqnarray*}

Therefore, for $0\le\epsilon\le\frac{1}{2}$,
\begin{eqnarray*}
\frac{1}{2}\Big\|
|\Psi\rangle\langle\Psi|
-|\Psi'\rangle\langle\Psi'|
\Big\|_1
&=&\sqrt{1-|\langle\Psi|\Psi'\rangle|^2}\\
&\le&\sqrt{1-(1-2\epsilon)^2}\\
&=&\sqrt{4\epsilon-4\epsilon^2}.
\end{eqnarray*}

\section{Correctability of errors}
\label{sec:fujii}
Let us consider an $n$-qubit graph state $|G\rangle$
and a tensor product $P$ of $n$ Pauli
operators. When $P$ acts on $|G\rangle$,
an $X$ operator
in $P$ can always be changed into the tensor product
of nearest-neighbour $Z$ operators by using the stabilizer relation. 
Therefore, we can 
always find $u\in\{0,1\}^n$ such that 
\begin{eqnarray*}
P|G\rangle=\Big(\bigotimes_{j=1}^nZ_j^{u_j}\Big)|G\rangle
=|G_u\rangle~(\mbox{up to phase}).
\end{eqnarray*}
Hence let us consider only $Z$ errors and error
is specified by $u\in\{0,1\}^n$.
Let $\{ \mathcal{M}_{a}\}_a$ be a POVM
corresponding to a fault-tolerant measurement-based quantum computation,
where $a$ is the output of the computation.
Then if 
\begin{eqnarray*}
{\rm Tr} [\mathcal{M}_a (| G\rangle \langle G| )] = 
{\rm Tr} [\mathcal{M}_a (| G_u\rangle \langle G_u| )] 
\end{eqnarray*}
for all $a$, $u$ is called a correctable error. 
The conditions, $Cond_1$ and $Cond_2$, are given as
the sets of syndromes on $|G_u\rangle$ for all correctable errors $u$.
An explicit form of the POVM depends on 
the fault-tolerant scheme chosen,
and therefore so does the set of correctable errors.
Most fault-tolerant schemes in the measurement-based model
are constructed by (or at least
can be regarded as) simulation of
circuit-based fault-tolerant schemes.
For example, fault-tolerant schemes 
in Ref.~\cite{RHG} and Refs.~\cite{DHN1,DHN2}
can be viewed as circuit-based fault-tolerant schemes 
using the Steane 7-qubit code and the surface code, respectively.
In the fault-tolerant theory for the circuit model,
a set of sparse errors are defined such that they do not change 
the output of the quantum computation under fault-tolerant quantum 
error correction~\cite{BA}.
Therefore it is straightforward to find a 
correctable set $\Gamma$ of errors by directly translating 
the set of sparse errors in the existing 
circuit-based fault-tolerant schemes 
into errors on the graph state in the measurement-based model. 
A channel ${\mathcal E}_n$ is not too strong so that
errors are correctable with high probability
if 
\begin{eqnarray*}
\mbox{Tr}\Big(\sum_{u\in \Gamma}|G_u\rangle\langle G_u| 
{\mathcal E}_n(|G\rangle\langle G|)\Big)\ge 1-\delta.
\end{eqnarray*}
Here, $\delta=2^{-poly(n)}$ for natural noises.
(In this paper, the proof holds even for sufficiently small
constant $\delta$.)
According to the theory of fault-tolerant quantum computation,
under a natural physical assumption like spatial locality
of noise, if noise strength of each noisy operation is sufficiently
smaller than a certain threshold value, the above condition
is satisfied~\cite{RHG,DHN1,DHN2,BA,AL}.

\acknowledgements
TM is supported by
Grant-in-Aid for Scientific Research on Innovative Areas
No.15H00850 of MEXT Japan, and the Grant-in-Aid
for Young Scientists (B) No.26730003 of JSPS.
KF is supported by KAKENHI No.16H02211.
HN is supported by the Grant-in-Aid for Scientific
Research (A) Nos.26247016 and 16H01705 of JSPS,
the Grant-in-Aid for Scientific Research
on Innovative Areas No. 24106009 of MEXT,
and the Grant-in-Aid for Scientific
Research (C) No.16K00015 of JSPS.

\end{document}